\documentclass[12pt,preprint2]{aastex6}

\usepackage{amsmath}
\usepackage{natbib}
\usepackage{rotating}

\newcommand{\nii}{\rm{[\ion{N}{2}]$\lambda$6584}}
\newcommand{\oiii}{\rm{[\ion{O}{3}]$\lambda$5007}}
\newcommand{\oiiii}{\rm{[\ion{O}{3}]$\lambda\lambda$4959,5007}}
\newcommand{\oii}{\rm{[\ion{O}{2}]$\lambda$3727}}
\newcommand{\ha}{\rm H{$\alpha$}}
\newcommand{\hb}{\rm H{$\beta$}}
\newcommand{\hii}{\rm{\ion{H}{2}}}
\newcommand{\siii}{\rm{[\ion{S}{2}]$\lambda\lambda$6717,6731}}
\newcommand{\oi}{\rm[\ion{O}{1}]$\lambda$6300}
\newcommand{\gmass}{\log(M_*/M_{\sun})}

\slugcomment{}        

\shorttitle{}
\shortauthors{F. Bian et al.}

\begin{document}


\title{Mass-Metallicity Relation for Local Analogs of High-Redshift galaxies:
Implications for the Evolution of the Mass-Metallicity Relations}


\author{Fuyan Bian\altaffilmark{1,6}, Lisa J. Kewley\altaffilmark{1}, Michael A.
Dopita\altaffilmark{1,2}, Guillermo A. Blanc\altaffilmark{3,4,5}}
\altaffiltext{1}{Research School of Astronomy \& Astrophysics, Mt Stromlo
Observatory, Australian National University, Canberra, ACT, 2611, Australia}
\altaffiltext{2}{Department of Astronomy, King Abdulaziz University, P.O. Box
80203 Jeddah, Saudi Arabia}
\altaffiltext{3}{Departamento de Astronom�a, Universidad de Chile, Camino del
Observatorio 1515, Las Condes, Santiago, Chile}
\altaffiltext{4}{Centro de Astrofísica y Tecnologías Afines (CATA), Camino del
Observatorio 1515, Las Condes, Santiago, Chile}
\altaffiltext{5}{Visiting Astronomer, Observatories of the Carnegie Institution
for Science, 813 Santa Barbara St, Pasadena, CA, 91101, USA}
\altaffiltext{6}{Stromlo Fellow}


\begin{abstract}
We revisit the evolution of the mass-metallicity relation of low- and
high-redshift galaxies by using a sample of local analogs of high-redshift
galaxies. These analogs share the same location of the UV-selected star-forming
galaxies at $z\sim2$ on the {\oiii/\hb} versus {\nii/\ha} nebular emission-line
diagnostic (or BPT) diagram. Their physical properties closely resemble those in
$z\sim2$ UV-selected star-forming galaxies being characterized in particular by high
ionization parameters ($\log q\approx7.9$) and high electron densities
($n_e\approx100~\rm{cm}^{-3}$). With the full set of well-detected rest-frame
optical diagnostic lines, we measure the gas-phase oxygen abundance in the SDSS
galaxies and these local analogs using the empirical relations and the photoionization models.
We find that the metallicity difference between the SDSS galaxies and our local
analogs in the $8.5<\gmass<9.0$ stellar mass bin varies from -0.09 to 0.39 dex,
depending on strong-line metallicity measurement methods. Due to this
discrepancy the evolution of mass-metallicity should be used to compare with the
cosmological simulations with caution. We use the [\ion{S}{2}]/H$\alpha$ and
[\ion{O}{1}]/H$\alpha$ BPT diagram to reduce the potential AGN and shock
contamination in our local analogs. We find that the AGN/shock influences are
negligible on the metallicity estimation.
\end{abstract}



\section{Introduction}
The dependence of chemical abundances on galaxy properties across cosmic time
provides insight into the physical mechanisms regulating the formation and
evolution of galaxies
\citep[e.g.,][]{Finlator:2008fj,Lilly:2013aa,Dave:2012aa,Ma:2016aa,Lu:2015aa}.
Correlations between stellar masses of galaxies and their gas-phase oxygen
abundances are well established in nearby galaxies
\citep{Tremonti:2004aa,Savaglio:2005aa,Kewley:2008aa,Andrews:2013aa,
Gonzalez-Delgado:2014aa}. Heavy elements that are expelled into the interstellar
medium (ISM) by supernovae explosions and stellar winds increase the
metallicity of galaxies when their stellar mass is built up. Galaxies with
higher stellar masses tend to have higher metallicity than lower mass galaxies. 
A strong evolution of the mass-metallicity relation has been
claimed in high-redshift studies
\citep[][]{Erb:2006rt,Maiolino:2008lr,Zahid:2013aa,Zahid:2014aa,Ly:2014aa,
Ly:2016ab,Maier:2014aa,Steidel:2014aa,Sanders:2015aa,Salim:2015aa,Guo:2016aa}. 

These studies typically use strong nebular emission-line ratios to estimate the
gas-phase oxygen abundance based on either photoionization models
\citep[e.g.,][]{Kewley:2002fk} or empirical calibrations \citep[e.g.,][PP04
hereafter]{Pettini:2004qe}. In particular, high-redshift metallicity
measurements rely heavily on the PP04 empirical calibrations. PP04 compiled an
extensive sample of \ion{H}{2} regions in nearby spiral, irregular and blue
compact galaxies and fit the relationship between the metallicities mostly from
direct temperature ($T_e$) method\footnote{PP04 also used photoionization models to measure
the metallicities for a small fraction of high metallicity objects in their
sample.} and strong emission-line ratios, including the $N2 = \log$(\nii/\ha)
and $O3N2 =\log$[(\oiii/\hb)/(\nii/\ha)] ratios. These two empirical calibrations
based on local \ion{H}{2} regions have been widely used to estimate the
metallicities of high-redshift galaxies
\citep[e.g.,][]{Erb:2006rt,Hainline:2009aa,Bian:2010vn,Steidel:2014aa,
Sanders:2015aa}. 

The interstellar medium (ISM) conditions, including the ionization parameter and
the ISM pressure, in high-redshift galaxies are quite different from those in
low-redshift galaxies
\citep[e.g.][]{Kewley:2013ab,Kewley:2013aa,Steidel:2014aa}. The ISM conditions
in high-redshift star-forming galaxies are characterized by $\sim0.6$~dex higher
ionization parameters and an order of magnitude higher ISM pressures/electron
densities than their local counterparts \citep[e.g.,][hereafter
B16]{Kewley:2013ab,Kewley:2013aa,Kewley:2015ab,Kewley:2016aa,Nakajima:2014aa,
Shirazi:2014ab,Sanders:2016aa,Dopita:2016aa,Bian:2016aa}. This evolution of the
ISM conditions raises questions on the applicability of local metallicity
calibrations for high redshift galaxies:  Are the empirical calibrations based
on the local \ion{H}{2} regions still valid for high-redshift galaxies
considering the dramatic changes of ISM conditions? How would the high
ionization parameters and the high ISM pressure in high-redshift galaxies affect
the metallicity estimation based on the $N2$ or $O3N2$ diagnostics?

Photoionization models provide important tools to investigate these questions.
Photoionization models can be used to establish the relations between
metallicity and certain emission line ratios, such as $R23 =
$[([\ion{O}{2}]$\lambda$3727+[\ion{O}{3}]$\lambda\lambda$4959,5007)/H$\beta$],
depending on the ISM conditions. Therefore, the detailed ISM conditions (e.g., ionization parameter,
electron density) are required as inputs for photoionization models. It is thus crucial to obtain high
signal-to-noise ratio (S/N) rest-frame optical spectra to detect lines, such as
the [\ion{O}{2}]$\lambda$3727, {\oiii}, and [\ion{S}{2}]$\lambda\lambda$6717,6731 emission lines. 
However, the spectra of high-redshift galaxies usually suffer from low S/N ($<5$) and limited wavelength
coverage. It is difficult to measure the ionization parameter and the electron
density in a large sample of individual high-redshift galaxies. 

In this paper, we use a sample of local analogs of high-redshift galaxies
selected based on their locations on the {\oiii/\hb} versus {\nii/\ha}
``Baldwin, Phillips \& Terlevich'' \citep[BPT,][]{Baldwin:1981rr} diagram.
These local analogs have about the same
properties as high-redshift galaxies, in particular high ionization parameters
and high electron densities. Therefore, these analogs provide local laboratories
to study the extreme star formation and the ISM conditions in high-redshift
galaxies. We use the well-determined physical conditions of ISM in our analogs
and photoionization models to study how the metallicities derived from strong
line measurements change with the ISM conditions and the implication for the
mass-metallicity relation measurements of high-redshift galaxies. The paper is
organized as follows. In Section~\ref{sec:sample}, we describe the selection of
the local analogs of high-redshift galaxies and the SDSS reference galaxy sample
for the further mass-metallicity relation study. In Section~\ref{sec:method}, we
describe the methods of metallicity measurement used in this paper. In
Section~\ref{sec:mzr}, we study the mass-metallicity relation in the local
analogs and the SDSS reference galaxy sample based on the eight metallicity
diagnostics. In Section~\ref{sec:discussion}, we discuss how AGNs/strong shocks
and the photoionization models affect the metallicity measurements in our local
analogs and high-redshift galaxies. In Section~\ref{sec:conclusions}, we
summarize the main conclusions of this paper.

Throughout this paper, we adopt following notations for the diagnostic emission
line flux ratios:
\begin{equation*}
N2=\log ({\rm [NII] \lambda6584/H\alpha})
\end{equation*}

\begin{equation*}
O3N2= \log[{\rm ([OIII]\lambda5007/H\beta)/[NII] \lambda6584/H\alpha}]
\end{equation*}

\begin{equation*}
N2O2=\log ({\rm [NII] \lambda6584/[OII]\lambda3727})
\end{equation*}

\begin{equation*}
R23 = {\rm ([OII]\lambda3727+[OIII]\lambda\lambda4959,5007)/H\beta}
\end{equation*}

\begin{equation*}
O32 = {\rm [OIII]\lambda\lambda4959,5007/[OII]\lambda3727}
\end{equation*}

The emission line flux ratios in this work are dust-extinction corrected by adopting
 \citet{Cardelli:1989aa} dust extinction law and assuming case B recombination
\citep[${\rm H\alpha/H\beta}=2.86$ for $T_e=10^4$ K][]{Osterbrock:2006oq}.

\begin{figure*}[]
\begin{center}
\includegraphics[scale=0.7,angle=-90]{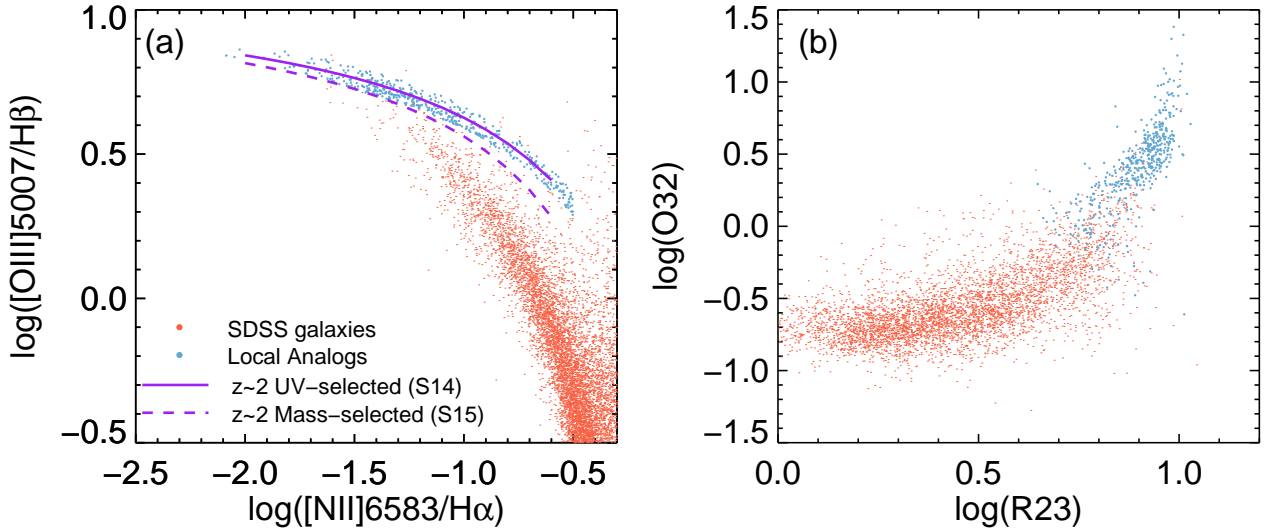}
\caption{Panel ({\it a}): [\ion{O}{3}]/{\hb} versus [\ion{N}{2}]/{\ha} BPT diagram.
The blue points represent the local analogs of high-redshift galaxies selected in
this work, and the red points represent the galaxies in the SDSS reference galaxy 
sample. The purple solid and dashed line represent the $z\sim2$ BPT locus in UV-selected
star-forming galaxies \citep[][S14]{Steidel:2014aa} and mass-selected star-forming galaxies
\citep[][S15]{Shapley:2015aa} at $z\sim2$, respectively. 
Panel ({\it b}): O32 versus R23 diagram. The blue points represent the local 
analogs of high-redshift galaxies and the red points represent the galaxies in
the SDSS reference galaxy sample.
\label{fig_select}}
\label{default}
\end{center}
\end{figure*}

\section{Sample Selection}\label{sec:sample}
\subsection{Local Analogs of High-redshift Galaxies}\label{analog}
We use a sample of local analogs of high-redshift galaxies to study the
metallicity estimation in high-redshift galaxies. We select local analogs
of high-redshift galaxies using the method developed in B16. 
B16 have demonstrated that the local galaxies selected based on the BPT diagram
share the same ISM conditions as star-forming galaxies at $z\sim2$.

In this study,
we select a sample of local analogs of high-redshift galaxy located in the
$\pm0.04$~dex region of the $z\sim2.3$ star-forming sequence defined by
equation~9 in \citet{Steidel:2014aa} on the {\oiii/\hb} versus {\nii/\ha}
BPT diagram (Figure~\ref{fig_select}(a))
\begin{equation}
\log({\rm[OIII]/H\beta)}>\frac{0.67}{\log({\rm[NII]/H\alpha)}-0.33}+1.09
\label{hz_lower},
\end{equation}
\begin{equation}
\log({\rm[OIII]/H\beta)}<\frac{0.67}{\log({\rm[NII]/H\alpha)}-0.33}+1.17
\label{hz_upper},
\end{equation} 
and
\begin{equation}
\log({\rm[NII]/H\alpha)} < -0.5\label{hz_right}
\end{equation}

A total of 458 unique galaxies are selected as local analogs of high-redshift galaxies
based on the above selection criteria (blue points in Figure~\ref{fig_select}(a)). 

Figure~\ref{fig_select}(b) shows the distribution of our analogs on the O32 versus 
R23 diagnostic diagram. These analogs also share the same region as high-redshift star-forming galaxies 
on this diagnostic diagram \citep{Nakajima:2013ab,Nakajima:2014aa,
Ly:2015aa,Ly:2016ab,Shapley:2015aa}, implying that these analogs share 
similar ionization parameters and metallicities with high-redshift star-forming
galaxies.

Here, we summarize the properties of the analogs. The
median stellar mass of our local analog is $\gmass=9.0\substack{+0.6 \\ -0.8} $ \footnote{In this paragraph, 
the uncertainties represent the 16th and 84th percentiles of the distribution of the parameters.}. The median SFR and sSFR
are $3.5\substack{+12.0\\-3.1}$~$M_{\sun}$~yr$^{-1}$ and $6.3\substack{+15.0\\-5.7}$~Gyr$^{-1}$, 
respectively. The sSFR of the
local analogs is comparable to that in $z\sim2$ star-forming galaxies with
similar stellar mass \citep[e.g.,][]{Rodighiero:2011fk}. Furthermore, these
analogs closely resemble the ISM conditions in $z\sim2-3$ galaxies \citep[e.g.,][]{Nakajima:2014aa,Sanders:2016aa}, 
including high ionization parameters ($\log q=7.9\pm0.2$~cm$^{-1}$) and high electron
densities ($n_e=120\substack{+146\\-106}$~cm$^{-3}$).

The local analog galaxies selected by the above criterion share the same location
of the BPT diagram with UV-selected galaxies $z\sim2$. UV-selected galaxies only
represent about 50\% of the full star-forming galaxy census at $z\sim2$
\citep[e.g.,][]{Reddy:2005aa,Ly:2011aa,Guo:2012aa}. \citet{Shapley:2015aa} found
that the [\ion{O}{3}]/{\hb} ratios in mass-selected star-forming galaxies 
are lower than those in UV-selected galaxies at $z\sim2$ \citep[also see][]{Dickey:2016aa}. 
Therefore, the ISM conditions
in our local analogs resemble those in $z\sim2$ UV-selected galaxies, but
may not fully represent the ISM conditions in all $z\sim2$ star-forming
galaxies.

\subsection{SDSS Reference Galaxy Sample}\label{SDSS}
We select a sample of galaxies observed in the SDSS from the MPA-JHU value added
catalog for SDSS Data Release 7 \citep[DR7,][]{Abazajian:2009aa}. The following
criteria are applied: (1) The objects are classified as either star-forming or
starburst galaxies in the MPA/JHU catalog, in which the criteria adopted from
\citet{Kewley:2001aa} were used to separate star-forming/starburst galaxies from
AGNs on the BPT diagram; (2) The S/Ns of [\ion{O}{2}]$\lambda$3727, H$\beta$,
[\ion{O}{3}]$\lambda\lambda4959, 5007$, H$\alpha$ and [\ion{N}{2}]$\lambda$6584
emission lines are greater than 10; (3) The fiber covering factors\footnote{The
ratio of the fiber flux to the total flux in $r$ band} are greater than 25\%. A
total of 91,469 SDSS galaxies that meet above criteria are selected. We refer
this sample of galaxies as ``SDSS reference galaxy sample" in this paper. 

In our local analog selections, we apply a $N2<-0.5$
cut to reduce the contamination of AGN. This selection criterion removes
metal-rich galaxies at the high mass end (see details in B16). We further apply
the same $N2$ cut to the SDSS galaxy and select a total of 39,875 galaxies. We
refer this sample of galaxies as ``SDSS reference galaxy sample with
$N2<-0.5$".

\section{Metallicity Measurements}\label{sec:method}
We implement the following eight strong line methods to measure the gas-phase
oxygen abundance in the local analogs of high-redshift galaxies and SDSS
galaxies in the reference sample. 

(1) We measure the gas-phase oxygen abundance using the $N2$- and
$O3N2$-metallicity relations adopted from PP04. PP04 established these empirical
relations by combining the direct $T_e$ metallicity and metallicity derived from
detailed photoionization modeling at high metallicity end
($12+\log(\rm{O/H}>8.5$) in a sample of {\hii} regions in nearby galaxies. The
$N2$ and $O3N2$ indices have been widely used to measure metallicities in high
redshift galaxies
\citep[e.g.,][]{Erb:2006rt,Hainline:2009aa,Bian:2010vn,Steidel:2014aa,
Sanders:2015aa}, even though it is unclear whether these empirical calibrations
are valid for high-redshift metallicity estimation due to the strong evolution
of the ISM conditions. 

(2) We measure the metallicities and the ionization parameters using the $R23$
and $O32$ indices by adopting the \citet[KK04 hereafter]{Kobulnicky:2004aa} recipe. 
KK04 established analytic relations
of $R23$, $O32$, the ionization parameter, and the metallicity by fitting the
\citet{Kewley:2001aa} photoionization models. These models use the STARBURST99
\citep{Leitherer:1999aa} stellar population synthesis models to generate the
input ionizing radiation field. This information is then inputted into the MAPPINGS III code \citep{Binette:1985aa,
Sutherland:1993aa} to construct photoionization models over a wide range
of the metallicities and the ionization parameters. We use the $N2O2$ to determine whether
each galaxy locates on the upper or lower $R23$ branch and compute the
metallicity and the ionization parameter iteratively until the metallicity
converge. We refer readers to section A2.3 in \citet{Kewley:2008aa} for more
detail. 
 
(3) We compute the metallicities and the ionization parameters simultaneously by
fitting multiple diagnostic lines with photoionization models using the IZI
program \citep[Inferring metallicities (Z) and Ionization
parameters,][]{Blanc:2015aa}. This program uses Bayesian inference to compute
the joint and marginalized probability density functions of the metallicity and
the ionization parameter for a set of emission line flux measurements based on
photoionization models. We use a set of metallicity and ionization parameter
sensitive emission lines, including {\oii}, {\hb}, {\oiiii}, {\ha}, {\nii}, and
{\siii} for the fitting and adopt the \citet{Kewley:2001aa} photoionization
models.

(4) We measure the metallicities using the new metallicity calibrations proposed
by \citet[D16 hereafter]{Dopita:2016aa}. This new calibration is based on a new grid of photoionization models from
the photoionization code MAPPINGS VI (Sutherland et al. 2015 in prep.). D16 found that the combination of {\nii/\siii}
and {\nii/\ha} provides a good metallicity diagnostics, which is not sensitive
to ionization parameters and electron densities. Furthermore, these lines are
close together in wavelength,  so they are less affected by reddening and can be
easily observed simultaneously in high-redshift galaxies. 

(5) We adopt the metallicities in the MPA/JHU catalog from \citet[][T04
hereafter]{Tremonti:2004aa}. The metallicity for each galaxy is estimate by
fitting the \citet{Charlot:2001aa} models to strong emission lines, including
{\oii}, {\hb}, {\oiiii}, {\ha}, {\nii}, and {\siii}. The \citet{Charlot:2001aa}
models use the \citet{Bruzual:2003lr} stellar synthesis models to characterize
the ionizing radiation field and the photoionization code CLOUDY
\citep{Ferland:1998aa} is applied to compute the emission line ratios.

(6) We compute the metallicity using the HII-CHI-mistry
\citep[HCm,][]{Perez-Montero:2014aa}. This code fits a large grids of
photoionization models to the emission line ratios of {\oii}/{\hb},
{\oiii}/{\hb}, {\nii}/{\hb}, and {\siii}/{\hb}. The photoionization models are
computed using the photoionization code CLOUDY \citep{Ferland:2013aa} and the
ionizing radiation field generated from the POPSTAR \citep{Molla:2009aa}.

(7) We compute the metallicity using the $R23$ and $O32$ indices from
\citet[][M91 hereafter]{McGaugh:1991aa}.  The M91 calibration is based on the
detailed \ion{H}{2} region models using the photoionization code CLOUDY
\citep{Ferland:1998aa} and the \citet{Mihalas:1972aa} stellar atmosphere models.

The methods (2)-(7) are all based on photoionization models codes, either
MAPPINGS or CLOUDY. These methods either solve the relations between the
metallicity and a set of emission line ratios (e.g., $R23$, $O32$) analytically
(e.g., KK04, D16, M91) or derive the metallicity by fitting a full set of
emission-line relative flux to photoionization grids (e.g., IZI, T04, HCm).

\begin{figure*}[]
\begin{center}
\includegraphics[width=2.2\columnwidth,angle=-90]{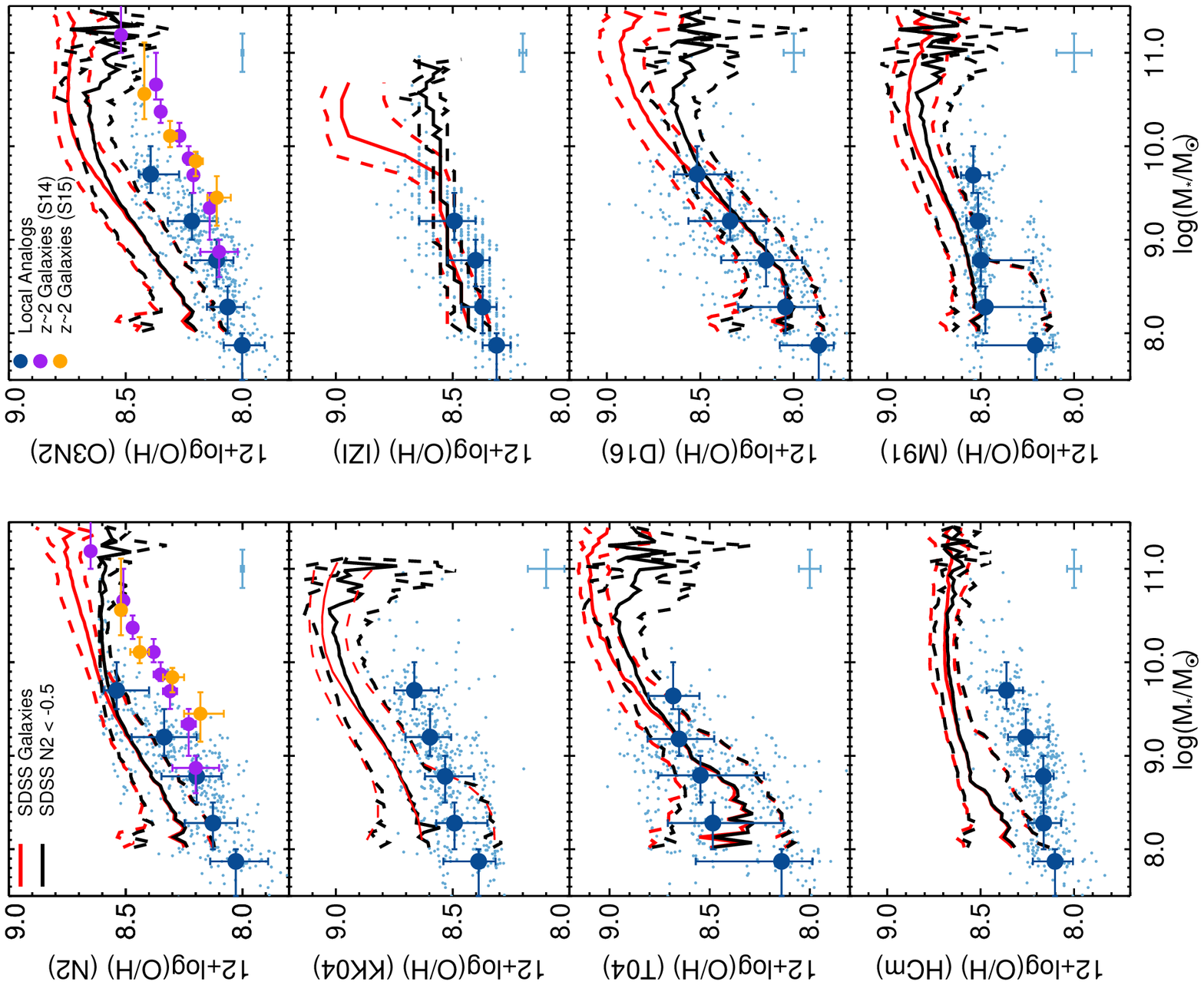}
\caption{Mass-metallicity relation of SDSS galaxies and the local analogs of
high-redshift galaxies. The metallicities are estimated based on $N2$ and $O3N2$
indicators by \citet{Pettini:2004qe}, $R23$ and $O32$ indicators by
\citet[][KK04]{Kobulnicky:2004aa} and \citet[M91]{McGaugh:1991aa}, combination
of $N2$ and $N2S2$ indicators by \citet[][D16]{Dopita:2016aa}, and fitting
photoionization models with all the available diagnostic lines simultaneously by
IZI code \citep{Blanc:2015aa}, \citet[][T04]{Tremonti:2004aa}, and HCm  code
\citep[][]{Perez-Montero:2014aa}. The light blue points represent the
mass-metallicity relation of individual local analogs, and the dark blue points
with error bars represent the median metallicity of local analogs in each
stellar mass bin. The horizontal error bar shows the stellar mass range in each
mass bin, and the vertical error bar represents the 16th and 84th percentiles of
the oxygen abundance distribution in each mass bin.  The red solid line
represents the median mass-metallicity relation of SDSS galaxies, and two dashed
lines represent the 16th and 84th percentiles of the oxygen abundance
distribution. The black solid curves represent the SDSS galaxies with $N2<-0.5$
cuts as we applied for our local analog selection. The error bar at the right bottom
corner of each of the plot represents a typical uncertainty of metallicity and
stellar mass estimation for an individual local analog galaxy. 
The purple points and orange
triangles with error bars represent the mass-metallicity relation of $z\sim2$
star-forming galaxies from \citet[S14]{Steidel:2014aa} and
\citet[S15]{Sanders:2015aa}, respectively. The horizontal error bar shows the
stellar mass range in each mass bin, and the vertical error bar indicates the
uncertainty in oxygen abundance estimated from the uncertainty on composite
emission-line fluxes for each stellar mass bin.
\label{fig_massmetal}}
\end{center}
\end{figure*}

\section{Mass-Metallicity Relation}\label{sec:mzr}

We compare the mass-metallicity relations of the SDSS galaxies in the reference
sample and our local analogs of high-redshift galaxies for each metallicity
calibration. Figure~\ref{fig_massmetal} shows the evolution of mass-metallicity
relations between the SDSS reference galaxies (red lines) and our local analogs (blue
circles). We quantify the evolution by computing the metallicity
difference between the SDSS galaxies and the local analogs ($\Delta \rm{Z} =
Z(SDSS)-Z(analog)$) for a given galaxy stellar mass bin.

We carefully select the appropriate mass bin to use for the metallicity
comparison. First, we choose a mass bin that is not affected by
the selection effect on the $N2 < -0.5$ cut. We compare the mass-metallicity
relation in the SDSS main galaxy sample and SDSS galaxy sample with $N2<-0.5$
and find that the two mass-metallicity relations
are consistent with each other at the low mass end and start to show
discrepancies at $\gmass>9.5$ in all eight diagnostic methods 
(red and black dashed lines in Figure~\ref{fig_massmetal}). Therefore, the
selection effect ($N2<-0.5$) does not affect the mass-metallicity relation at
$\gmass<9.5$. Secondly, we would like to choose the mass bin whose median
metallicity of the local analogs are comparable to that in $z\sim2-3$ galaxies
(Figure~\ref{fig_massmetal}) . We choose the mass bin of $8.5<\gmass<9.0$ to
compare the $\Delta \rm{Z}$ between local analogs and SDSS galaxies to meet the
above two requirements. 

Table~\ref{tab.result} summarizes the metallicity differences between the SDSS
galaxies and the local analogs ($\Delta \rm{Z}$) in the mass bin of
$8.5<\gmass<9.0$ based on the eight metallicity estimation methods in
Figure~\ref{fig_massmetal}. There exist significant metallicity
differences between the SDSS galaxies and the local analogs in all eight
metallicity estimation methods. The $\Delta \rm{Z}$ changes from -0.09~dex
to 0.39~dex depending on the method used.

The $\Delta \rm{Z}$ results derived from the two empirical calibrations (N2
and O3N2) are not consistent with each other. $\Delta \rm{Z}(O3N2)$ is larger
than $\Delta \rm{Z}(N2)$ by about 0.05~dex. This discrepancy was also found in
high-redshift mass-metallicity relation studies
\citep[e.g.,][]{Steidel:2014aa,Zahid:2014ac,Sanders:2015aa}. This discrepancy is
primarily caused by the offset between the local galaxies and high-redshift
galaxies on the {\oiii/\hb} versus {\nii/\ha} BPT diagram. This offset
could be due to higher electron densities and ionization parameters, harder ionization radiation
fields, various N/O ratios, AGN/shock contributions, and/or selection effects
\citep[e.g.,][]{Liu:2008kx,Brinchmann:2008ab,Kewley:2013ab,Kewley:2013aa,Juneau:2014aa,Masters:2014aa,
Masters:2016aa,Newman:2014aa,Steidel:2014aa,Shapley:2015aa,Bian:2016aa}.

The $\Delta \rm{Z}$ results derived from the photoionization model grids
are different. The KK04, IZI and D16 calibrations are all based on the MAPPINGs
photoionization models, however, the $\Delta \rm{Z}$s from these three methods
vary from $-0.03$ to $0.19$. The D16 and IZI methods yield a much smaller $\Delta
\rm{Z}$ than the KK04 calibration (Table~\ref{tab.result}). The major cause of
this discrepancy is that the R23 methods have a fixed metallicity (e.g., O2N2
value) threshold to break the lower and upper $R23$ branch degeneracy. However,
the R23 metallicity upper/lower branch turnover points decrease with the
increasing ionization parameters in photoionization models \citep[e.g., Figure~7
in][]{Kobulnicky:2004aa}. This metallicity threshold is chosen to meet the
conditions of low ionization parameter in local star-forming galaxies. However,
this threshold tends to place galaxies with high ionization parameters on the
upper $R23$ branch onto the lower $R23$ branch, which underestimates the
metallicity in our analog galaxies and overestimates the $\Delta \rm{Z}$. On the
other hand, the IZI method considers all the available emission line
information, providing a more robust metallicity estimation based on the
photoionization models. 

The ionizing radiation field also makes significant impact on the $\Delta
\rm{Z}$ results based on the photoionization models. For example, both T04 and
HCm methods fit a full set of emission-line relative fluxes to the
photoionization models from CLOUDY code, however, the $\Delta \rm{Z}$s show
$\sim0.4$~dex difference. One of the major differences of these two
photoionization models is that the input ionizing radiation field, that T04 and
HCm implement the BC03 and POPSTAR stellar synthesis models, respectively. The
BC03 model yields harder UV spectra than the POPSTAR model \citep{Molla:2009aa}.

\begin{table*}
\centering
\caption{The metallicity difference between SDSS galaxies and local analogs of
high-redshift galaxies in $8.5<\gmass<9.0$ mass bin\label{tab.result}}
\begin{tabular}{lcccccccc}
\hline
Method & N2 & O3N2 & KK04 & IZI & T04 & D16 & HCm & M91   \\
\hline
\hline
 $\Delta$Z \footnote{The first row: the metallicity difference between the SDSS
galaxies in the reference sample and the full sample of local analogs.} & $0.19\substack{+0.11\\-0.15}$ \footnote{The errors represent the 16th and 84th percentiles 
 of the $\Delta$Z distribution.}&
$0.25\substack{+0.07\\-0.11}$ & $0.19\substack{+0.12\\-0.08}$ & $0.12\substack{+0.06\\-0.12}$ & $-0.09\substack{+0.32\\-0.21}$ & $-0.03\substack{+0.19\\-0.24}$ 
& $0.39\substack{+0.06\\-0.10}$ & $0.07\substack{+0.28\\-0.04}$  \\
$\Delta$Z \footnote{The second row: the metallicity difference between the SDSS
galaxies in the reference sample and local analogs without potential AGN/shock
contamination.}  &$0.18\substack{+0.11\\-0.14}$&$0.24\substack{+0.07\\-0.11}$&$0.18\pm0.07$&$0.12\substack{+0.06\\-0.12}$
&$-0.12\substack{+0.28\\-0.19}$&$-0.07\substack{+0.09\\-0.19}$
&$0.40\substack{+0.04\\-0.10}$&$0.06\substack{+0.05\\-0.04}$\\
\hline
\end{tabular}
\end{table*}

\begin{figure*}[]
\begin{center}
\includegraphics[scale=0.7,angle=-90]{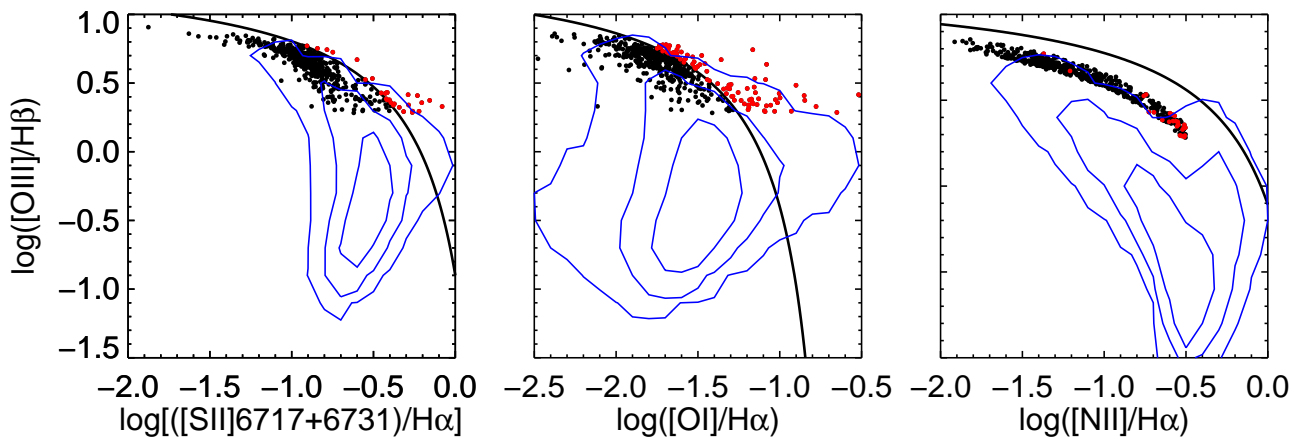}
\caption{S2, O1, and N2 BPT diagrams. The blue contours represent 1$\sigma$, 2$\sigma$, and 3$\sigma$ distributions of the SDSS reference galaxy sample in the BPT diagrams. The black solid line is the separation of
star-forming region and AGN/shock region in the BPT diagram
\citep{Kewley:2006aa}. The red filled circles represent the local analog
galaxies with potential AGN/shock contamination, and the black circles represent
the local analog galaxies with pure star formation.\label{fig_bpts2o2}}
\label{default}
\end{center}
\end{figure*}

\begin{figure*}[]
\begin{center}
\includegraphics[width=2.2\columnwidth,angle=-90]{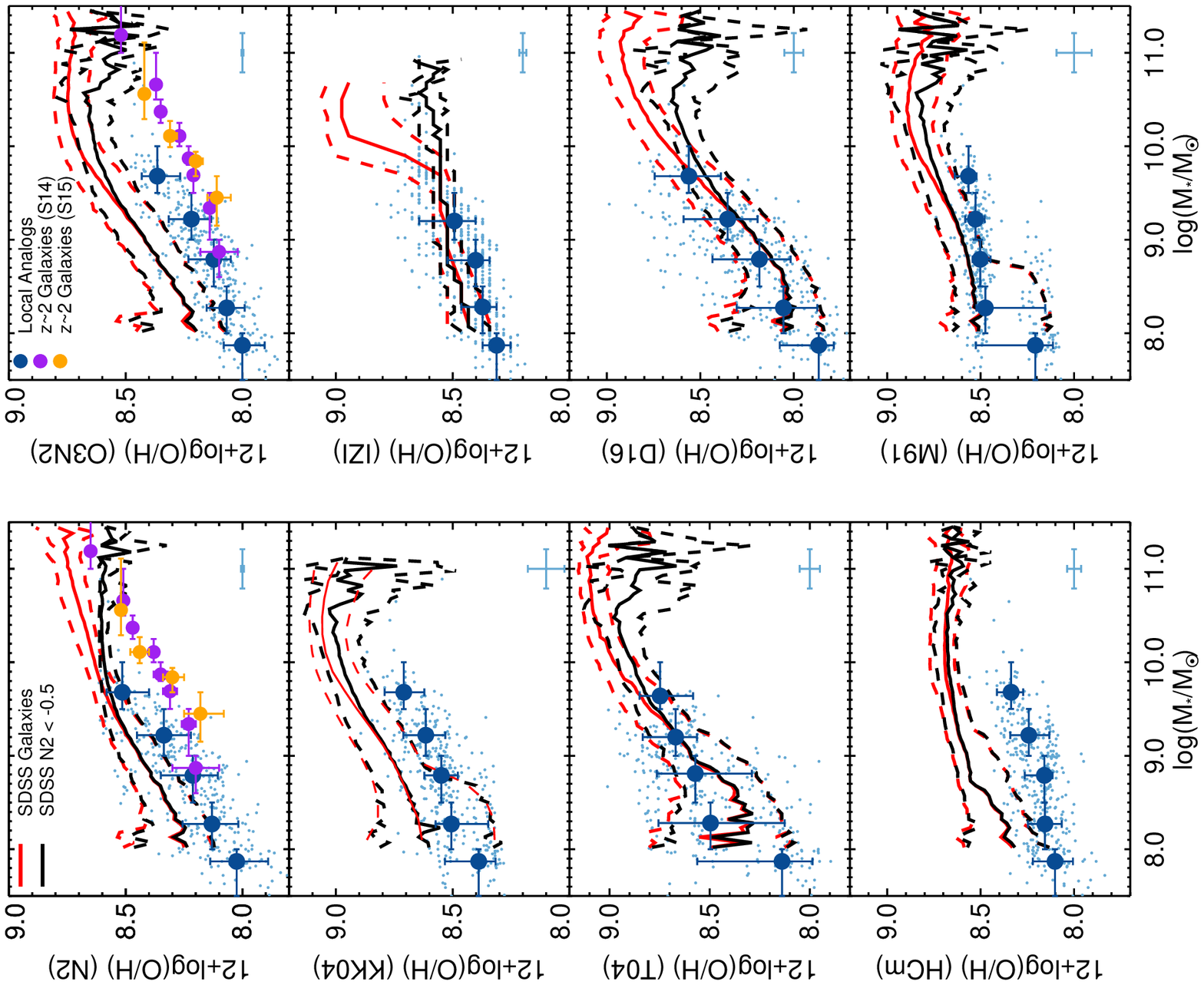}
\caption{Similar to Figure~1, but we exclude local analog galaxies
that are potentially affected by AGNs and/or shocks based on the S2 and O1 BPT diagrams.
AGN/shock contamination does not have a significant impact on the mass-metallicity relation.
\label{fig_massmetal_cut}}
\end{center}
\end{figure*}

\section{Discussion}\label{sec:discussion}
\subsection{Effects of Shocks and AGNs}
Shocks and/or AGNs in galaxies could enhance the {\nii} flux, resulting in an
overestimation of the $N2$-based metallicity for galaxies, in particular in
high-redshift galaxies \citep[e.g.,][]{Newman:2013aa,Maier:2014aa}. The local
analogs allow us to access the diagnostic lines that are sensitive to the shocks
and/or AGNs, including the {\siii}, {\oi} emission lines. These lines are
sensitive to the hardness of the ionizing radiation field. We use the
[\ion{O}{3}]/H$\beta$ versus [\ion{S}{2}]/H$\alpha$ and [\ion{O}{1}]/H$\alpha$
BPT diagnostic diagrams (S2 and O1 BPT diagrams) to remove the local analogs
whose {\nii} emission line flux could be potentially contaminated by AGN/shock
excitation. 

Figure~\ref{fig_bpts2o2} shows the S2 and O1 BPT diagnostic diagrams. The solid
lines are adopted from \citet{Kewley:2006aa} to separate the pure star forming
galaxies and galaxies with AGNs/strong shocks. We offset the
\citet{Kewley:2006aa} criteria by +0.05 dex in the [\ion{O}{3}]/H$\beta$
direction to meet the conditions in high-redshift galaxies. In the BPT diagrams,
we are able to identify galaxies that are potentially contaminated by
AGNs/strong shocks (red squares in Figure~\ref{fig_bpts2o2}). We establish a
new clean sample of the local analogs of high-redshift galaxies by removing
galaxies ($\sim$20\% of the total sample of the local analogs) located in the
AGN/shocks regions either in the S2 or O1 BPT diagram. The
[\ion{N}{2}]/H$\alpha$ in this new sample of local analogs is less affected by
AGNs and/or strong shocks. We thus expect that the $N2$-based metallicity is
more reliable in this new clean sample. Figure~\ref{fig_massmetal_cut} shows the
mass-metallicity relation in the new clean sample of local analogs of
high-redshift galaxies, and we also summarize the new $\Delta \rm{Z}$ results
estimated from the eight methods in the second row of Table~\ref{tab.result}. We
find that mass-metallicity relation in this new sample of the local analogs is
consistent with that in the original sample of local analogs. It suggests that
shocks/AGNs do not significantly affect the $N2$-based metallicity estimation in
the local analogs of high-redshift galaxies. The local analogs assemble similar
physical properties to high-redshift star-forming galaxies (B16). We thus expect
the effects of AGN/strong shocks would be negligible for the global metallicity
measurement of high-redshift galaxies, in particular at the low mass end
($\gmass<9.5$).

\subsection{Applicability of the Photoionization Models}
The metallicity estimation based on photoionization models relies on the
accuracy of the photoionization models. The systematic uncertainties in
photoionization models could affect the metallicity estimation in galaxies
\citep[e.g.,][]{Kewley:2002fk,Blanc:2015aa}. The two major uncertainties are the
input ionizing radiation field based on the stellar synthesis models and the
star formation history, and the relation between N/O and oxygen abundance (O/H).

The harder ionizing radiation field provides more high energy ionizing photons,
which enhance the fluxes of the {\nii} and {\oiii} emission line. Therefore, the
harder radiation field increases the $N2$, $O3N2$, $R23$ indices for a given
metallicity, affecting the metallicity estimation based on the above metallicity
diagnostic indices. \citet{Steidel:2014aa} proposed that high-redshift galaxies
have a harder radiation field due to the stellar binarity and rotation
\citep[e.g.,][]{Eldridge:2009aa,Levesque:2010aa,Steidel:2016aa}. This requires
the application of the photoionization models with different ionizing radiation
field to low- and high-redshift galaxies.

The $N2$ and $O3N2$ indices are also sensitive to nitrogen abundance, so the
relation between N/O and oxygen abundance is crucial for obtaining a reliable
oxygen abundance measurements.  An evolution in the N/O versus O/H  relation with
redshift \citep[e.g.,][]{Masters:2014aa, Masters:2016aa,Shapley:2015aa} would
affect the metallicity estimates based on the $N2$ and $O3N2$ indices
\citep[c.f.][]{Dopita:2016aa,Strom:2016aa}. This evolution does not affect the
$R23$ index, but it can cause issues when using the $N2O2$ index to determine
whether one galaxy is located in the upper or lower $R23$ branch. 

However, we have very limited information on how the radiation field and N/O
evolve with redshift to provide reliable inputs for the photoionization models.
Further detailed studies on either deep spectra of the individual analog
galaxies or composite spectra of a sample of local analogs would provide us
better constraints on the evolution of radiation field and N/O ratio (Bian et
al. in prep).

\section{Conclusions}\label{sec:conclusions}
We investigate the mass-metallicity relation in the local analogs of
high-redshift galaxies. These local analogs have similar interstellar medium
(ISM) conditions to high-redshift galaxies, which provide a great opportunity to
understand how the ISM conditions affect the metallicity measurements in
high-redshift galaxies. We summarize our main results as follows:
\begin{itemize}
\item The well-detected full set of rest-frame optical diagnostic lines allow us
to study the mass-metallicity relation in the local analogs and normal SDSS
galaxies in the reference sample using eight different metallicity
diagnostics, including both empirical calibrations and metallicity diagnostics
based on photoionization models.

\item The metallicity difference between the local analogs and normal SDSS
galaxies in the mass bin of $8.5<\gmass<9.0$ varies between -0.09 and 0.39
dex depending on the strong-line metallicity diagnostic method used.

\item We remove the local analogs whose emission lines are potentially
contaminated by AGN/shock excitation using the [\ion{O}{3}]/H$\beta$ versus
[\ion{S}{2}]/H$\alpha$ and [\ion{O}{1}]/H$\alpha$ BPT diagnostic diagrams. The
mass-metallicity relation does not change after we remove the local analogs with
AGN/shocks contamination, suggesting that the AGN/shock contribution in our
local analogs is negligible for the global metallicity estimation in our local
analogs.
\end{itemize}

\acknowledgments
We are very grateful to the anonymous referee for comments and suggests to
improve the manuscript. M.D. and L.K, acknowledge the support of the Australian
Research Council (ARC) through Discovery project DP130103925. M.D. would also
like to thank the Deanship of Scientific Research (DSR), King AbdulAziz
University for additional financial support as Distinguished Visiting Professor
under the KAU Hi-Ci program. G.B. is supported by CONICYT/FONDECYT, Programa de
Iniciacion, Folio 11150220.



\facility{SDSS}

\bibliography{paper}

\clearpage





\end{document}